\documentclass[12pt]{article} 
\usepackage{graphicx,floatflt,amssymb,rotate}
\usepackage{mathrsfs}
\usepackage{amssymb}
\usepackage{amsmath}
\usepackage{color}
\usepackage{graphicx}
\usepackage{subfigure}
\usepackage{multirow}
\usepackage{pstricks}
\usepackage{float}
\usepackage[toc,page]{appendix}

\begin{document}
\vskip 30pt  
 
\begin{center}  
{\Large \bf A note on gauge-fixing in the electroweak sector of nmUED} \\
\vspace*{1cm}  
\renewcommand{\thefootnote}{\fnsymbol{footnote}}  
{ {\sf Anindya Datta$^1$ \footnote{email: adphys@caluniv.ac.in}},  
{\sf Avirup Shaw$^2$ \footnote{email: avirup.cu@gmail.com}} 
}\\  
\vspace{10pt}  
{ {\em $^1$Department of Physics, University of Calcutta,  
92 Acharya Prafulla Chandra Road, \\ Kolkata 700009, India}
\\
{\em $^2$Regional Centre for Accelerator-based Particle Physics, Harish-Chandra Research Institute,
\\ Chhatnag Road, Jhusi, Allahabad 211019, India}}
\normalsize  
\normalsize   
\end{center} 

%\pacs{11.10.Kk}{Field theories in dimensions other than four}
%\pacs{11.15.-q}{Gauge field theories}
%\pacs{12.60.-i}{Models beyond the standard model}

\begin{abstract}
Electroweak observables are highly sensitive to the loop corrections. Therefore, a proper gauge-fixing mechanism is always needed to define the propagators which are involved in Feynman loop amplitude. With this spirit we compute gauge-fixing mechanism in five dimensional (5-D) Universal Extra-Dimensional (UED) model with boundary localised terms (BLTs). These BLTs  are not 5-D operators in four-dimensional (4-D) effective
theory but some sort of boundary conditions on the respective fields at the
fixed points of $S^1/Z_2$ orbifold. Furthermore, these BLTs non-trivially modify the Kaluza-Klein (KK) spectra and
some of the interactions among the KK-excitations compared to the minimal UED (mUED), in which, these BLTs are absent. In this note we calculate the gauge-fixing mechanism in the electroweak sector of such non-trivial UED scenario. Moreover, we discuss the composition and masses of Goldstone and any physical scalar that emerge after the symmetry breaking in this set up with different choices of gauge.
\end{abstract}

\noindent PACS No: {\tt 11.10 Kk, 11.15.-q, 12.60.-i}\renewcommand{\thesection}{\Roman{section}}  
\setcounter{footnote}{0}  
\renewcommand{\thefootnote}{\arabic{footnote}}

%\maketitle

%% here a revision

%\revision{Insert here the text.
%See fig.~\ref{fig.1}, table~\ref{tab.1} and eq.~(\ref{eq.1}).
%See also~\cite{b.a,b.b}.}

%here a shortcut $\emc$ and again $\emc$

%\begin{equation}
%\label{eq.1}
%0\neq1
%\end{equation}

%\begin{figure}
%\onefigure{epl-template.eps}
%\caption{Figure caption.}
%\label{fig.1}
%\end{figure}

%\begin{table}
%\caption{Table caption.}
%\label{tab.1}
%\begin{center}
%\begin{tabular}{lcr}
%first  & table & row\\
%second & table & row
%\end{tabular}
%\end{center}
%\end{table}

%\section{Section title }
Extra dimensional models in which all the Standard Model (SM) fields propagate in one or more compactified space-like extra dimensions offer viable solutions to several long standing problems of Particle Physics  like Dark Matter \cite{ued_dm1, ued_dm2, ued_dm3, ued_dm4, ued_dm5}, Unification \cite{ued_uni1, ued_uni2, ued_uni3} etc. We are interested in a particular incarnation of such models namely UED \cite{acd} where all the SM fields can feel a single compactified extra space-like dimension $y$. The effective 4-D action of UED, consists of SM particles and towers of their KK-excitations. Members of a tower is specified by so called 
KK-number ($n$), which is nothing but the discretised momentum in direction of the extra dimension. Masses of the KK-excitations in $n^{th}$ KK-state is typically $nR^{-1}$, where $R$ is the radius of compactification. 
The interactions among the fields are governed by the $SU(3)_C \times SU(2)_L \times U(1)_Y$ gauge symmetry of the SM. To accommodate the chiral fermions of the SM, one has to demand for some extra $Z_2$ symmetry of the action after orbifolding the extra direction. These results into two special points 
$y =0$ and $y = \pi R$ called the {\em fixed points}, along the $y$ direction. 

Radiative corrections to the masses of the KK-states are important from phenomenological view point and has been considered in \cite{rad_cor_georgi, rad_cor_cheng}. UED being an effective theory, lacks the ultraviolet completion in radiative corrections.  So assumptions have been made in Ref.\cite{rad_cor_cheng}, such that at the cut-off scale, radiative corrections to all KK-masses identically vanish. 
In principle, one can take into account the contributions (to the radiative corrections) beyond the cut-off, by adding BLTs at the fixed points ($y =0$ and $y = \pi R$) with their coefficients as free parameters. These models are collectively called non-minimal UED (nmUED). Theoretical as well as different possibilities of phenomenological aspects of nmUED have been investigated to some extent in existing literature \cite{nmUED1, nmUED2, nmUED3, nmUED4, carena, schwin, flacke, ddrs, ashes1, ashes2, ddrs_dm, drs, as, zbb, bsmm, Dey:2016cve}.

In this note, we discuss the gauge-fixing procedure in the context of nmUED model. In literature there exist several articles \cite{uedgf1, uedgf2, uedgf3, uedgf4} in which the issue of gauge-fixing procedure has been discussed in the context of UED, but in case of nmUED it would not be a straight forward job due to the presence of boundary terms. To our knowledge, the first time we are going to analyse this issue in electroweak sector of this non-trivial extension of UED scenario. Furthermore, we will discuss the masses and composition of Goldstone and other physical scalars emerging after the electroweak symmetry breaking. At this point we would like to make a comment that the gauge-fixing mechanism for 5-D QED in presence of BLTs, has been discussed in \cite{Muckphd}.  We extend the prescription in Ref.\cite{Muckphd} to the case when  the symmetry is spontaneously broken in a {\it non-abelian gauge theory}.

Let us start with the free action (governed by $SU(2)_L \times U(1)_Y$ gauge group) of 5-D gauge and scalar fields with their respective boundary localised kinetic terms (BLKTs) \cite{flacke, ddrs, ashes1, ashes2, ddrs_dm, drs, as, zbb, bsmm}: 

\begin{eqnarray}
S_V &=& -\frac{1}{4}\int d^5x \bigg[ W_{MN} W^{MN}+  r_W \left\{ \delta(y) +  \delta(y - \pi R)\right\} W_{\mu\nu} W^{\mu \nu}\nonumber \\
&&+B_{MN} B^{MN}+ r_B \left\{ \delta(y) +  \delta(y - \pi R)\right\} B_{\mu\nu} B^{\mu \nu}\bigg],
\label{pure-gauge}
\end{eqnarray}

\begin{eqnarray} 
S_H &=& \int d^5x \bigg[ (D_{M}H)^\dagger(D^{M}H)+ r_H \left\{ \delta(y) +  \delta(y - \pi R)\right\} (D_{\mu}H)^\dagger(D^{\mu}H) \bigg].
\label{higgs}
\end{eqnarray}

Here, $r_W$, $r_B$ and $r_H$ parametrise the strength of the BLKTs for the respective fields. 5-D field strength tensors are given below:
\begin{eqnarray}\label{ugfs}
W_{MN}^a &\equiv& (\partial_M W_N^a - \partial_N W_M^a-{\tilde{g}_2}\epsilon^{abc}W_M^bW_N^c),\\ \nonumber
B_{MN}&\equiv& (\partial_M B_N - \partial_N B_M).
\end{eqnarray}
$W^a_M$ and $B_M$ ($M=0,1 \ldots 4$) are the 5-D gauge fields corresponding to $SU(2)_L$ and $U(1)_Y$ gauge group respectively. 5-D covariant derivative is defined as $D_M\equiv\partial_M+i{\tilde{g}_2}\frac{\sigma^{a}}{2}W_M^{a}+i{\tilde{g}_1}\frac{Y}{2}B_M$, with the 5-D gauge coupling constants ${\tilde{g}_2}$ and ${\tilde{g}_1}$. ${\sigma^{a}}\over 2$ and $\frac Y2$ are the corresponding generators of the gauge groups. $H=\left(\begin{array}{cc}
\phi^+\\\phi^0\end{array}\right)$ is the 5-D Higgs doublet with $\phi^0=\frac{(\tilde{v}+h+i\chi)}{\sqrt{2}}$, $\tilde{v}$ being the 5-D vacuum expectation value (VEV).

Before going to discuss gauge-fixing mechanism we would like to make some clarifying remarks which could help the reader to understand the following {\it gauge-fixing} action. First of all, generally KK-decomposition of neutral gauge bosons become very complicated by the fact that $B$ and $W^3$ mix in the bulk as well as on the boundary. So, unless $r_W=r_B$, it would not be possible to diagonalise the bulk and boundary actions simultaneously by the same 5-D field redefinition. In the following we will stick to the $r_W=r_B$ equality condition. However, in general one can deal with $r_W\neq r_B$, but in this case the mixing term between $B$ and $W^3$ in the bulk and on the boundary points generate off-diagonal terms in the neutral gauge boson mass matrix. In this situation Unitary gauge (in which it is difficult to calculate loop corrections) choice is the only one favourable option where we can play with this nmUED scenario \cite{ddrs, ashes1, ashes2, ddrs_dm, drs, as}. The reason is due to the fact that in UED or nmUED like scenario one can argue about the Unitary gauge choice condition which is equivalent to large $R^{-1}$ limit \cite{ddrs, ashes1, ashes2, ddrs_dm, drs, as}. As in this case contribution to the KK-masses from spontaneous breaking of the electroweak symmetry are negligible with respect to extra-dimensional contribution. Later we will discuss this issue elaborately.

Eq.\ref{pure-gauge} and \ref{higgs} must be supplemented by the gauge-fixing action.  One of the aims for fixing the gauge  is to get rid of the terms in which $V_\mu$ couples to $V_4$ (arise in Eq.\ref{pure-gauge}) and/or to $\eta (\equiv\phi^+~{\rm or}~\chi)$ (in Eq.\ref{higgs}). Here we use $V$ to generically denote vector fields and also we use $r_V$ as the generic BLKT parameter for gauge bosons. Further, we use $\eta$ as the generic notation for 5-D Goldstone fields. To ensure the (second) cancellation, Higgs BLKT parameter $r_H$ should be included in the gauge-fixing action. We have adapted the gauge-fixing action from \cite{Muck:2001yv, Muck:2002er}, where a similar kind of gauge-fixing action can be found in presence of boundary loclaised terms. However, we have considered the following gauge-fixing action appropriate for nmUED model\footnote{From now and rest of our analysis we use $\partial_4 (D_4)\equiv\partial_y (D_y)$.} (for the purpose of writing gauge-fixing action we use the conventional notations $W^\pm,~Z$ (for $weak~gauge~bosons$) and $A$ (for $photon$), however in rest of our analysis we will use generic notation $V$),

\begin{eqnarray} 
S_{GF} &=& -\frac{1}{\xi _y}\int d^5x\Big\vert\partial_{\mu}W^{\mu +}+\xi_{y}(\partial_{y}W^{4+}+iM_{W}\phi^{+}\{1 + r_{H}\left( \delta(y) + \delta(y - \pi R)\right)\})\Big \vert ^2 \nonumber \\&&-\frac{1}{2\xi_y}\int d^5x [\partial_{\mu}Z^{\mu}+\xi_y(\partial_{y}Z^{4}-M_{Z}\chi\{1+ r_H( \delta(y) +  \delta(y - \pi R))\})]^2
\nonumber \\&&-\frac{1}{2\xi_y}\int d^5x [\partial_{\mu}A^{\mu}+\xi_y\partial_{y}A^{4}]^2. 
\label{gauge-fix}
\end{eqnarray}\\

%with $M_V = \frac{1}{2}g_{\rm i} v\sqrt{\frac{1+\frac{r_V}{\pi R}}{1+\frac{r_H}{\pi R}}}$.  

The above gauge-fixing action is somewhat very special and at the same time very crucial for this nmUED scenario. Due to the presence of the BLKTs in the Lagrangian lead to a non-homogeneous weight function for the fields with respect to the extra dimension. This inhomogeneity forces us to choose a $y$-dependent gauge-fixing parameter $\xi_y$ as \cite{Muckphd},

\begin{equation}
\xi =\xi_y\,(1+ r_H\{ \delta(y) +  \delta(y - \pi R)\}),
\end{equation}
here $\xi$ is independent of $y$. This relation can be seen as {\em renormalisation} of the gauge-fixing parameter as the 
BLKTs are in some sense counter terms taking into account the unknown ultraviolet contribution in loop calculations. In this sense, 
$\xi_y$ is the bare gauge-fixing parameter while $\xi$ can be viewed as the renormalized gauge-fixing parameter taking the values $0$ (Landau gauge),
$1$ (Feynman gauge) or $\infty$ (Unitary gauge).

Appropriate KK-expansion of the fields which are involved in the above actions can be schematically written as:
\begin{equation}\label{Amu}
V_{\mu}(x,y)=\sum_n V_{\mu}^{(n)}(x) a^n(y)
\end{equation}
\begin{equation}\label{A4}
V_{4}(x,y)=\sum_n V_{4}^{(n)}(x) b^n(y),
\end{equation}
and
\begin{equation}\label{chi}
H(x,y)=\sum_n H^{(n)}(x) f^n(y).
\end{equation}

Variation of the action functional equations (Eq.\ref{pure-gauge} and \ref{higgs}) and utilizing Eq.\ref{Amu}, \ref{A4} and \ref{chi} we obtain equation for the $y$-dependent wave functions of $V_\mu$ as follows \cite{flacke, ddrs}:

\begin{eqnarray}
\bigg[ \partial^2 _y + m_{V^n}^2+\{ \delta(y) +  \delta(y - \pi R)\}\{r_V m_{V^n}^2 -(r_H-r_V)M_V^2\}\bigg] a^n (y) = 0.
\label{gauge equation}
\end{eqnarray}\\
The above equation can be used for both $W$ and $Z$ bosons. $M_V$ is the electroweak symmetry breaking mass. For, $f^n (y)$ and $b^n (y)$ we have \cite{carena, ddrs}:
\begin{equation}
\left[ \partial^2 _y + m_{H^n}^2 (1 + r_H\{ \delta(y) +  \delta(y - \pi R)\})\right] F^n (y) = 0,
\label{Higgs equation}
\end{equation}
where $F^n (y)\equiv f^n (y)$ or $b^n(y)$. Here, $m_{V^n}(m_{H^n})$ are the KK-mass solutions of gauge (Higgs) fields obtained from transcendental equation which we will discuss in the following.  

For the sake of completeness in Eq.\ref{gauge equation} we have shown the equation of motion for $a^n(y)$. However, the Eq.\ref{gauge equation} or its solution will not play any major role in our analysis. One can see that in case of~~$r_V = r_H$,  Eq.\ref{gauge equation} has the identical form as that of Eq.\ref{Higgs equation} and they have the same solutions. In the following we will show without loss of any generality that this {\it equality} condition helps us to fix the gauge properly in this nmUED scenario.

We must keep in mind that the zero mode fields must  correspond to respective SM particles.  However, some of the fields in 5-D theory do not have any SM counterpart. Consequently application of appropriate boundary conditions, while solving the differential equation (Eq.\ref{Higgs equation}), is very crucial. As for example, $V_{\mu}(x,y)$ and $H(x,y)$ should have zero  modes, while $V_{4}(x,y)$ cannot have a zero mode. 

Using  the proper boundary conditions \cite{carena, ddrs}, solutions  for $f^n(y)$ can  be written in a compact form\cite{flacke, zbb, bsmm} as $N_{n} \cos[ m_{H^{n}} (y-\frac{\pi R}{2})]$ for even (including the zero mode) KK-modes and $N_{n} \sin [m_{H^{n}} (y - \frac{\pi R}{2})]$ for odd KK-modes. Here $N$'s are the normalisation constant. Note that these functions have a definite parity of $(-1)^n$ (called KK-parity) under 
$y \leftrightarrow y - \pi R$ transformation\footnote{This is equivalent to a reflection symmetry along a line $y = \frac{\pi R}{2}$ 
in the $y$-space.}.

The eigenvalues $m_{H^n}$'s are the solutions of the transcendental equation \cite{carena,ddrs_dm, zbb, bsmm}:
 
\begin{eqnarray}
  \frac{r_{H} m_{H^{n}}}{2}= \left\{ \begin{array}{rl}
         -\tan \left(\frac{m_{H^{n}}\pi R}{2}\right) &\mbox{for $n$ even,}\\
          \cot \left(\frac{m_{H^{n}}\pi R}{2}\right) &\mbox{for $n$ odd.}
          \end{array} \right.   
          \label{tranc}      
\end{eqnarray}

It should be noted that zero mode of $V_4$ is not desirable, 
one can easily write the  solutions of $y$-dependent part of  $V_{4}(x,y)$ i.e., $b^n(y)$ as $N_n\sin [m_{H^{n}} (y - \frac{\pi R}{2})]$ and $N_n\cos [m_{H^{n}} (y-\frac{\pi R}{2})]$ for even and odd KK-modes respectively.

From the above discussions it is evident that $y$-dependent wave functions $b^n(y)$ and $f^n(y)$ are transformed in opposite way under $Z_2$ parity, consequently $b^n(y)$ and $f^n(y)$ should satisfy the relation $\partial_{y} f^n (y) = m_{H^n} b^{n}(y)$. 
In fact, $f^n$'s are so called {\em primitive}
 of $b^n$'s \cite{Muckphd}. In general a function $\tilde b^n(y)$ would be called the primitive of  $b^n
(y)$, if 
\begin{equation}\label{primitive}
\partial_{y}\tilde{b}^n(y)=m_{H^n} b^{n}(y). 
\end{equation}
Thus, $b^{n}(y)$ and $\tilde{b}^n(y)$ obey the following condition \cite{Muckphd}:
\begin{equation}\label{odd1}
\frac{m_{H^n}\partial_{y}b^{n}(y)}{[1+ r_H\{ \delta(y) +  \delta(y - \pi R)\}]}=-m^2_{H^n}\tilde{b}^n(y),
\end{equation}
or equivalently we can write it as (using Eq.\ref{primitive}),
\begin{equation}\label{odd2}
[\partial^2_{y}+m^2_{H^n}(1+ r_H\{ \delta(y) +  \delta(y - \pi R)\})]\tilde{b}^n(y)=0.
\end{equation}
This equation (equivalent to Eq.\ref{Higgs equation}) immediately gives the following orthogonality relations \cite{ddrs}:
\begin{eqnarray}
&&\hspace*{-1cm}\int_0 ^{\pi R}\hspace*{-.5cm}
dy \; \left[1 + r_{f}\{ \delta(y) + \delta(y - \pi R)\}\right]\tilde{b}^n(y)\tilde{b}^m(y)=\delta^{n m},\label{orthonorm1}\\
&&\int_0 ^{\pi R}\hspace*{-.5cm}
dy \; \partial_{y}\tilde{b}^n(y)\partial_{y}\tilde{b}^m(y)
=m^2_{H^n}\delta^{n m}.
\label{orthonorm2}
\end{eqnarray}
In principle, $a^n (y)$ could also be a primitive of $b^n(y)$ when
 $r_H$ would equal to $r_V$. So in our analysis $\tilde{b}^n\equiv f^n,~a^n$.

Now let us come to the main point of our interest. What are the composition and masses of any physical scalar mode that 
emerges after the symmetry breaking? To find this out one has to look into the mass-mixing matrix constructed out of the $V^{(n)}_4~(\equiv W^{+(n)}_4~{\rm or}~Z^{(n)}_4) $
and the $\eta^{(n)}~(\equiv\phi^{+(n)}~{\rm or}~\chi^{(n)})$. All the elements of this matrix are obtained from Eq.\ref{higgs} and Eq.\ref{gauge-fix}. In Eq.\ref{higgs} we have a term like $(D_{y}H)^\dagger(D^{y}H)$ from which the relevant terms are as follows $\frac12 M^2_V V_4V^4$, $\frac12(\partial_y\eta)(\partial^y\eta)$ and $M_V(\partial_y\eta)V^4$. Using the KK-decompositions (Eq.\ref{A4} and Eq.\ref{chi}), primitive condition (Eq.\ref{primitive}) and orthogonality relation (Eq.\ref{orthonorm2}), we can express the first two terms as $-\frac12M^2_VV^{(n)}_4V^{(n)}_4$, $-\frac12m^2_{H^n}\eta^{(n)}\eta^{(n)}$ and these two terms contribute to diagonal elements of the mass-mixing matrix. Again using Eq.\ref{chi}, Eq.\ref{primitive} and Eq.\ref{orthonorm2} we can express the remaining term\footnote{As we have already mentioned that we generically use $\eta$ instead of $\phi^+$ and $\chi$, so for $\eta$ we can use the KK-decomposition given in Eq.\ref{chi}.} as $-M_Vm_{H^n}\eta^{(n)}V^{(n)}_4$ which contributes to off-diagonal elements. The complementary parts of the matrix elements (for both diagonal and off-diagonal) are generated from the Eq.\ref{gauge-fix} (gauge-fixing action) and the relevant terms (written in terms of generic notation $V$ and $\eta$) are given by $-\frac{\xi (\partial_y V^4)^2}{2(1+ r_H\{ \delta(y) +  \delta(y - \pi R)\})}$, $-\frac{\xi}{2} M^2_V (1+ r_H\{ \delta(y) +  \delta(y - \pi R)\})\eta^2$ and $\xi M_V(\partial_y V^4)\eta$. Then again using KK-decompositions (Eq.\ref{A4} and Eq.\ref{chi}), Eq.\ref{odd1} and Eq.\ref{orthonorm1} we can express these terms as $-\frac{\xi}{2}m^2_{H^n}V^{(n)}_4V^{(n)}_4$, $-\frac{\xi}{2}M^2_V\eta^{(n)}\eta^{(n)}$ and $\xi M_Vm_{H^n}\eta^{(n)}V^{(n)}_4$ respectively. Note that, the construction of mass-mixing matrix elements of the above forms are possible only in the limit $r_V=r_H$, as otherwise we can not identify $\tilde{b}^n\equiv f^n$. Using above expressions we can write the mass-mixing matrix in $V^{(n)}_4(x)  - \eta^{(n)}(x)$ basis in $R_{\xi}$ gauge as:

\begin{equation} 
M^n_{\xi}
 = 
\begin{pmatrix}
M_V^2+\xi m_{H^n}^2 & (1-\xi) M_V m_{H^n} \\
(1-\xi) M_V m_{H^n} & \xi M_V^2+m_{H^n}^2
\end{pmatrix}
.
\label{mass_matrix}
\end{equation}

Let us now pay some attention to the eigenvalues of this matrix $M^n_{\xi}$. A
straight forward diagonalisation would results into two eigenvalues
$\lambda _1 = (M_V^2 + m_{H^n}^2)$ and $\lambda _2 = \xi\;(M_V^2 + m_{H^n}^2)$ and corresponding eigenstates are given by:
\begin{equation}\label{phys}
{\mathcal{H}}^{(n)} = \frac{\left(m_{H^n}\eta^{(n)}-M_{V}V^{(n)}_4\right)}{M_{V^{(n)}}},
\end{equation}
\begin{equation}\label{gldsn}
{\mathcal{G}}^{(n)} = \frac{\left(M_{V}\eta^{(n)}+m_{H^n}V^{(n)}_4\right)}{M_{V^{(n)}}},
\end{equation}
where, $M_{V^{(n)}}=\sqrt{M_V^2 + m_{H^n}^2}$.
It is thus very simple to identify the state with mass(-square)
$\lambda_1$ with the physical scalar (as this is independent of gauge
choice) and the remaining one with the Goldstone.  Note that, in the
Feynman gauge (with $\xi = 1$) these two states are mass degenerate.
Unitary gauge is equivalent to choosing $\xi \rightarrow \infty$, when this second
state becomes infinitely heavy and decouples from the theory. 
And, finally in the Landau gauge (with $\xi = 0$), the $\xi$ dependent eigenvalue becomes zero, 
resulting in vanishing determinant of the mass-matrix and a massless Goldstone mode characteristic of this gauge. The mixing angle ($\theta$) is independent of the gauge choice as, $\tan 2 \theta = 
\frac{2 M_V m_{H^n}}{m_{H^n}^2 - M_V^2}$. The orthogonal combination to the Goldstone, with eigenvalue ($M_A^2 + m_{H^n}^2$)  is a physical 
scalar which remains in the theory. Mass of this physical mode is independent of the gauge choice. To this end let us remind that, though the structure of the orthogonal combination of the physical scalar and Goldstone are identical with UED scenario, but the extra dimensional contributions are obtained from the solution of transcendental equation (Eq.\ref{tranc}). Furthermore, without the equality condition $r_V = r_H$, such simplified version (identical with UED scenario) of gauge-fixing procedure would not be possible in nmUED scenario.

In literature  \cite{carena, ddrs, drs, as},  taking $V_4 \rightarrow 0$ is synonymous with the Unitary gauge. Here, we would
 like to  shed light on this issue in context of our earlier discussion. Extremising the gauge fixed action of $V_M$ (Eq.\ref{pure-gauge}
 and Eq.\ref{gauge-fix}) would lead to an equation of motion for $V_\mu (x, y)$ (under the assumption 
 of $r_V = r_H$):

\begin{eqnarray}
\hspace*{-1cm}\bigg[[\partial^{\mu}\partial_{\mu} g^{\mu\nu}+M_V^2g^{\mu\nu}-(1-\frac{1}{\xi})\partial^{\nu}\partial^{\mu}][1+r_V\{ \delta(y) +  \delta(y - \pi R)\}]-\partial_y^2g^{\mu\nu}\bigg]V_{\mu} (x, y)=0.
\label{D_equation}
\end{eqnarray}

Using $\xi \rightarrow \infty$ in the above equation would lead us to the equation of motion for $V_\mu$ in the Unitary gauge.
  However one can also achieve this without gauge-fixing and simply by setting $V_4 \rightarrow 0$ in Eq.\ref{pure-gauge} followed by extermisation of the action. In fact, setting $V_4$ to zero in Eq.\ref{pure-gauge}
  would  straightforwardly remove the undesirable terms in which $V_\mu$ couples to $V_4$ via derivative, which was the main purpose of gauge-fixing. And, as long as we are interested in $V_\mu$ and its interactions with other physical particles, taking $V_4 \rightarrow 0$ is as good as Unitary gauge\footnote{An equivalent statement can also be found in the Ref \cite{Muck:2001yv} in which all fermions are localised on the 4-D
brane.}. However we must keep in mind that the {\em Goldstone} is a linear combination of $V^{(n)}_4$ and $\eta^{(n)}$ (see Eq.\ref{gldsn}). And the coefficients of this linear combination is independent of choice of gauge. Whenever, one needs to have the interactions involving physical scalar (like Charged Higgs) and/or Goldstones (in some gauge other than the Unitary gauge), one should be more careful and should use the proper definitions of these scalars. For example, the Refs.\cite{zbb, bsmm, Dey:2016cve} deal with the electroweak radiative corrections in nmUED scenario and the entire calculations have been performed in Feynman gauge where Goldstone bosons appear in the loop diagrams. The prescription of gauge-fixing mechanism in the above Refs.\cite{zbb, bsmm, Dey:2016cve} has been adopted from the formalism presented in this letter. A more careful look at the expression for the mixing angle in the physical scalar/Goldstone sector would reveal that in the large $R^{-1}$ limit ($R^{-1} >> v({\rm SM~VEV})$),  physical scalar can be identified with $\eta$ (see Eq.\ref{phys}) while $V_4$  becomes the Goldstone (see Eq.\ref{gldsn}). 
  
In summary, we have discussed, the gauge-fixing procedure in a model of Universal Extra Dimensions with boundary localised kinetic terms where a gauge symmetry is spontaneously broken by Higgs mechanism.  Finally we have discussed the composition and masses of the Goldstone and physical scalar in this model.

{\bf Acknowledgements}
AD is partially supported by  Department of Atomic Energy-Board of Research in Nuclear Sciences (DAE-BRNS) research project. AS acknowledges financial support Department of Atomic Energy, Government of India, for the Regional Centre for Accelerator-based Particle
Physics (RECAPP), Harish-Chandra Research Institute. Authors are grateful  to Amitava Raychaudhuri for carefully reading the manuscript and several illuminating discussions.


\begin{thebibliography}{99}

\bibitem{ued_dm1}G. Servant and T. M. P. Tait, Nucl. Phys. B {\bf650} (2003) 391, [arXiv:hep-ph/0206071].

\bibitem{ued_dm2}D.~Majumdar, Mod.\ Phys.\ Lett.\ A {\bf 18} (2003) 1705.

\bibitem{ued_dm3}K. Kong and K. T. Matchev, JHEP {\bf 0601} (2006) 038 [arXiv:hep-ph/0509119].

\bibitem{ued_dm4}F.~Burnell and G.~D.~Kribs, Phys.\ Rev.\ D {\bf 73} (2006) 015001, [arXiv:hep-ph/0509118].

\bibitem{ued_dm5}G. Belanger, M. Kakizaki and A. Pukhov, JCAP {\bf 1102} (2011) 009, [arXiv:1012.2577 [hep-ph]].



\bibitem{ued_uni1} K. R. Dienes, E. Dudas and T. Gherghetta, Phys. Lett. B {\bf436} (1998) 55, [arXiv:hep-ph/9803466].
\bibitem{ued_uni2}K. Dienes, E. Dudas, and T. Gherghetta, Nucl. Phys. B {\bf537} (1999) 47, [arXiv:hep-ph/9806292].
\bibitem{ued_uni3}G. Bhattacharyya, A. Datta, S. K. Majee, A. Raychaudhuri, Nucl.Phys. B {\bf760} (2007) 117, [arXiv:hep-ph/0608208].

\bibitem{acd} 
T.~Appelquist, H.~C.~Cheng and B.~A.~Dobrescu, 
 %``Bounds on universal extra dimensions,'' 
 Phys.\ Rev.\ D {\bf 64} (2001) 035002, [arXiv:hep-ph/0012100]. 

\bibitem{rad_cor_georgi}H.~Georgi, A.~K.~Grant and G.~Hailu,
  %``Brane couplings from bulk loops,''
  Phys.\ Lett.\ B {\bf 506} (2001) 207,
  [arXiv:hep-ph/0012379].
\bibitem{rad_cor_cheng}H. C.~Cheng, K. T.~Matchev and M.~Schmaltz,  
%``Radiative corrections to Kaluza-Klein masses,''  
Phys.\ Rev.\ D {\bf 66} (2002) 036005,  
[arXiv:hep-ph/0204342]. 

\bibitem{nmUED1} G.~R.~Dvali, G.~Gabadadze, M.~Kolanovic and F.~Nitti,
  %``The Power of brane induced gravity,''
Phys.\ Rev.\ D {\bf 64} (2001) 084004, [arXiv:hep-ph/0102216].
 
\bibitem{nmUED2}F.~del Aguila, M.~Perez-Victoria and J.~Santiago,
  %``Some consequences of brane kinetic terms for bulk fermions,''
JHEP {\bf 0302} (2003) 051, [arXiv:hep-th/0302023].
\bibitem{nmUED3}F.~del Aguila, M.~Perez-Victoria and J.~Santiago,
  %``Physics of brane kinetic terms,''
  Acta Phys.\ Polon.\ B {\bf 34} (2003) 5511, [arXiv:hep-ph/0310353]. 
\bibitem{nmUED4}T. Flacke, K. Kong and S.C. Park, JHEP {\bf 05} (2013) 111 [arXiv:1303.0872 [hep-ph]].

\bibitem{carena}  
%\cite{Carena:2002me} 
%\bibitem{Carena:2002me} 
 M.~S.~Carena, T.~M.~P.~Tait and C.~E.~M.~Wagner, 
 %``Branes and orbifolds are opaque,'' 
 Acta Phys.\ Polon.\ B {\bf 33} (2002) 2355 
 [arXiv:hep-ph/0207056]. 
 %%CITATION = HEP-PH/0207056;%% 
\bibitem{schwin}
C. Schwinn, Phys. Rev. D {\bf 69} (2004) 116005, [hep-ph/0402118].
\bibitem{flacke}  
%\cite{Flacke:2008ne} 
%\bibitem{Flacke:2008ne} 
 T.~Flacke, A.~Menon and D.~J.~Phalen, 
 %``Non-minimal universal extra dimensions,'' 
 Phys.\ Rev.\ D {\bf 79} (2009) 056009 
 [arXiv:0811.1598 [hep-ph]]. 
 %%CITATION = ARXIV:0811.1598;%%%\cite{Schwinn:2004xa} 
\bibitem{ddrs}
A. Datta, U. K. Dey, A. Shaw, and A. Raychaudhuri, Phys.
Rev. D {\bf 87} (2013) 076002 [arXiv:1205.4334 [hep-ph]].

\bibitem{ashes1} A.~Datta, K.~Nishiwaki and S.~Niyogi, JHEP {\bf 1211} (2012) 154, [arXiv:1206.3987 [hep-ph]].

\bibitem{ashes2} A.~Datta, K.~Nishiwaki and S.~Niyogi, JHEP {\bf 1401} (2014) 104, [arXiv:1310.6994 [hep-ph]].

\bibitem{ddrs_dm}
A. Datta, U. K. Dey, A. Raychaudhuri, and A. Shaw, Phys.
Rev. D {\bf 88} (2013) 016011, [arXiv:1305.4507 [hep-ph]].

\bibitem{drs} A. Datta, A. Raychaudhuri, and A. Shaw, Phys. Lett. B {\bf730} (2014) 42 , [arXiv:1310.2021 [hep-ph]]. 

\bibitem{as}A. Shaw, Eur. Phys. J. C {\bf 75} (2015) 33, [arXiv:1405.3139 [hep-ph]].  

\bibitem{zbb}T. Jha and A. Datta, JHEP {\bf 1503} (2015) 012, [arXiv:1410.5098 [hep-ph]].

\bibitem{bsmm}A. Datta and A. Shaw, Phys. Rev. D {\bf 93} (2016) 055048, [arXiv:1506.08024 [hep-ph]].

%\cite{Dey:2016cve}
\bibitem{Dey:2016cve} 
  U.~K.~Dey and T.~Jha,
  %``Rare Top Decays in Minimal and Non-minimal Universal Extra Dimension,''
  [arXiv:1602.03286 [hep-ph]].
  %%CITATION = ARXIV:1602.03286;%%
  %2 citations counted in INSPIRE as of 25 Mar 2016



\bibitem{uedgf1}
J.~Papavassiliou and A.~Santamaria,  Phys. Rev. D~{\bf 63}~(2001)~125014, [arXiv:hep-ph/0102019]. 
\bibitem{uedgf2}F. J. Petriello, JHEP {\bf 0205} (2002) 003, [arXiv:hep-ph/0204067]. 
\bibitem{uedgf3}A. J.~Buras, M.~Spranger and A.~Weiler,
%``The impact of universal extra dimensions on the unitarity triangle and 
%rare K and B decays. ((U)),'' 
Nucl.\ Phys.\ B {\bf 660} (2003) 225, [arXiv:hep-ph/0212143]. 

\bibitem{uedgf4}C. Csaki, J. Hubisz and P. Meade [arXiv:hep-ph/0510275].

\bibitem{Muckphd} A. M\"{u}ck, Ph. D dissertation, University of W\"{u}rzburg, 2004.[http://www.physik.uni-wuerzburg.de/fileadmin/11030200/Dissertationen/mueck-diss.pdf]

\bibitem{Muck:2001yv} 
  A.~Muck, A.~Pilaftsis and R.~Ruckl,
  %``Minimal higher dimensional extensions of the standard model and electroweak observables,''
Phys.\ Rev.\ D {\bf 65} (2002) 085037, [arXiv:hep-ph/0110391].
%doi:10.1103/PhysRevD.65.085037.
%%CITATION = doi:10.1103/PhysRevD.65.085037;%%
  %104 citations counted in INSPIRE as of 05 Aug 2016

\bibitem{Muck:2002er} 
  A.~Muck, A.~Pilaftsis and R.~Ruckl, [arXiv:hep-ph/0203032].
  %``Electroweak constraints on minimal higher dimensional extensions of the standard model,''

%%CITATION = HEP-PH/0203032;%%
  %14 citations counted in INSPIRE as of 05 Aug 2016





\end{thebibliography}
\end{document}